\documentclass[11pt]{article}

\usepackage{amsmath}
\usepackage{amsthm}
\usepackage{amssymb}
\usepackage{adjustbox} 
\usepackage{algorithmic}
\usepackage{bbm}
\usepackage{color}
\usepackage{caption,subcaption}
\usepackage{dsfont} 
\usepackage{graphicx}
\usepackage{hyperref}
\usepackage{mathrsfs}
\usepackage[margin=1in]{geometry}
\usepackage{mathtools}
\usepackage{rotating}
\usepackage{tikz}
\usepackage{wrapfig}
\usepackage{setspace}
\usepackage{booktabs} 
\usepackage{accents} 
\usepackage{soul} 
\usepackage{relsize} 
\usepackage{lscape} 
\usepackage{xcolor} 
\setstcolor{red}
\usepackage{multirow}
\usepackage{booktabs,siunitx}
\usepackage[utf8]{inputenc}
\usepackage{float}
\restylefloat{table}
\usepackage{enumitem}
\newlist{Steps}{enumerate}{1}
\setlist[Steps,1]{label=Step~\arabic*.,leftmargin=*}

\newcommand{\indep}{\mathbin{\rotatebox[origin=c]{90}{$\models$}}}

\newcommand{\E}{\mathbb{E}}

\newcommand{\A}{\mathbf{A}}

\newcommand{\C}{\mathbf{C}}

\newcommand{\Y}{\mathbf{Y}}

\newcommand{\bO}{\mathbfcal{O}}

\newlength{\dhatheight}

\DeclareMathAlphabet\mathbfcal{OMS}{cmsy}{b}{n}

\usepackage{accents}

\newcommand{\redsout}{\bgroup\markoverwith{\textcolor{red}{\rule[0.5ex]{2pt}{0.4pt}}}\ULon}

\usepackage{tikz}

\newtheorem{defn}{Definition}
\newtheorem*{defn*}{Definition}

\newtheoremstyle{break}
  {\topsep}{\topsep}%
  {\itshape}{}%
  {\bfseries}{}%
  {\newline}{}%
\theoremstyle{break}

\usepackage[numbers,sort&compress]{natbib}

\title{Missing data in non-stationary multivariate time series from digital studies in Psychiatry}
\author{Xiaoxuan Cai, Charlotte R. Fowler, Li Zeng, Habiballah Rahimi Eichi, Dost Ongur,\\ Lisa Dixon, Justin T. Baker, Jukka-Pekka Onnela, Linda Valeri \\[7pt] Department of Statistics, The Ohio State University \\[7pt] {cai.1083@osu.edu}}
\date{}

\begin{document}
\maketitle
\doublespace
\abstract{Mobile technology (e.g., mobile phones and wearable devices) provides scalable methods for collecting physiological and behavioral biomarkers in patients' naturalistic settings, as well as opportunity for therapeutic advancements and scientific discoveries regarding the etiology of psychiatric illness. 
Continuous data collection through mobile devices generates highly complex data: entangled multivariate time series of outcome, exposure, and covariates. 
Missing data is a pervasive problem in biomedical and social science research, and Ecological Momentary Assessment (EMA) data in psychiatric research is no exception. 
However, the complex data structure of multivariate time series and non-stationary nature make missing data a major challenge for proper inference.
Additional historical information included in time series analyses exacerbates the issue of missing data and also introduces problem for confounding adjustment. 
The majority of existing imputation methods are either designed for stationary time series or longitudinal data with limited follow-up periods. 
The limited work on non-stationary time series either focuses on missing exogenous information or ignores the complex temporal dependent relationship among outcome, exposure and covariates. 
We propose a Monte Carlo Expectation-Maximization algorithm of the state space model (MCEM-SSM) to effectively handle missing data in non-stationary entangled multivariate time series. 
We demonstrates the method’s advantages over other widely used missing data imputation strategies through simulations of both stationary and non-stationary time series, subject to various missing mechanisms. 
Finally, we apply the MCEM-SSM to a multi-year smartphone observational study of bipolar and schizophrenia patients to investigate the association between digital social connectivity and negative mood. 
\\[1em]
}

\section{Introduction}
Mobile devices (e.g., mobile phones and wearable devices) have revolutionized the way we access information and provide convenient and scalable methods for collecting psychological and behavioral biomarkers continuously in patients’ naturalistic settings \citep{world2011mhealth,ben2017mobile}.
An important application of mobile technology is in psychiatric research to monitor patients' psychiatric symptoms, social interaction, life habits
, and other health-related conditions.
Intensive monitoring increases the number of observations to hundreds or even thousands, resulting in intensive longitudinal data or entangled multivariate time series, which greatly complicates research design, statistical analysis, and the handling of missing data.

Our research is motivated by the Bipolar Longitudinal Study (BLS), a multi-year observational smartphone study for bipolar and schizophrenia patients \citep{valeri2023digitalpsychiatry,fowler2022}. Severe Mental Illness (SMI), including schizophrenia, bipolar disorder, schizoaffective disorder, and related conditions, has become one of the major burdens of disease, affecting over 13 million individuals in the United States \citep{NSDUH2019}. Antipsychotics, albeit critical in controlling manic and depressive episodes, fall short in improving patient's quality of life \citep{lieberman2005effectiveness}. 
On the other hand, social support, in particularly the size of social network \citep{corrigan2004social,hendryx2009social}, has been demonstrated to be critical for promoting sustained improvements in symptoms and social functioning \citep{johnson1999social,pevalin2003social,corrigan2004social,hendryx2009social}. 
It is therefore of interest to monitor closely patients' social activity and evaluate its correlation with psychiatric symptoms.
In the Bipolar Longitudinal Study, passive data, such as GPS, phone use, accelerometer data, and anonymized basic information of phone call and text logs, were collected continuously without any necessary direct involvement from study participants. 
Active data, such as self-reported symptoms, behaviors, and experiences, also known as Ecological Momentary Assessment (EMA) \citep{wichers2011momentary}, were obtained through daily reports from study participants. 
Leveraging the richness of this smartphone longitudinal study, we wish to investigate the association between digital social connectivity and patient's self-reported negative mood, while controlling for other behavioral and environmental factors (e.g., physical activity captured by accelerometer data and weather temperature). 
Despite the BLS study being unique in its multiyear follow-up and wealth of data collected in the context of SMI, we anticipate that long-term close monitoring of patients will become prevalent in the future as mobile technology becomes more prevalent in clinical settings. 

Missing data is a significant issue in mobile device data, driven by long-term follow up and intensive measurements.
It can arise from several reasons, including device power loss, disconnection, or participants'  nonresponse.
We identify three key challenges associated with missing data in mobile device-based time series research.
First, history information (e.g., lagged outcomes) is often included as explanatory variables in time series analysis to control for auto-correlation, which significantly elevates the missing rate in analysis.
Second, discarding time points with incomplete records disrupts the temporal structure of the data, potentially distorting relationships among variables and introducing bias into statistical inference.
Finally, psychiatric patients exhibit significant heterogeneity due to varying disease subtype, pharmacological and behavioral treatments, and 
exhibit non-stationarity over the lengthy follow-up period due to personal recovery trajectory.
Non-stationarity, characterized by changes in the mean or variance over time, is often induced by the evolving treatment effect and covariates' influences.
N-of-1 study design is a patient-centered research design that is well suited for personalized mobile device studies or mental health research with high patient heterogeneity \citep{kumar2013mobile}. 
Proper missing data imputation in psychiatric research must account for both participant heterogeneity and non-stationarity to ensure  unbiased analysis.

Advancements in handling missing data have emerged in statistics, economics, and computer science in recent decades. 
However, available approaches do not address our issue of missing data in  response variable and explanatory variables in non-stationary time series. 
Complete case analysis eliminates incomplete records and reduces estimation efficiency, especially for data with a high missing rate \citep{white2010bias,little2019statistical}; in time series data, it additionally disrupts the temporal relationships among variables and introduces bias in estimation \citep{white2010bias,bashir2018handling}. 
Common imputation methods for cross-sectional and longitudinal data include mean imputation, last-observation-carried-forward (LOCF) imputation, and linear and spline interpolation. A considerable body of literature has cautioned against their use, as they often produce biased estimation even under missing completely at random (MCAR) \citep{honaker2010missing}.
Multiple imputation, widely recommended 
in longitudinal studies \citep{twisk2002attrition,spratt2010strategies}, relies on time-invariant models for imputation, making it unsuitable for non-stationary time series. 
Weighted estimation equations (WEEs) are another class of methods for handling missing data for static processes, however, they are not well-suited for non-stationary time series and pose additional challenges when propensity scores approach zero for datasets with a large number of time points  \citep{ibrahim2005missing}. 
Moving average techniques have been applied to univariate time series but fail to capture the relationships among entangled multiple time series. 
More recent machine-learning approaches for multivariate time series (e.g., current neural networks \citep{fortuin2020gp}, generative adversarial networks \citep{luo2018multivariate}), require stationary data generation processes and large independently and identically distributed (i.i.d) samples for training, and are thus unsuitable for non-stationary mobile device time series data.

Model-based maximum likelihood (ML) methods provide a principled approach to deal with missing data under missing completely at random (MCAR), missing at random (MAR) and missing not at random (MNAR) \citep{little2019statistical,follmann1995approximate,ibrahim2009missing}, and are widely applied to longitudinal studies using (generalized) linear mixed-effects  models \citep{follmann1995approximate,ibrahim2009missing} or generalized estimation equations \citep{fitzmaurice2000generalized,troxel1998analysis}. 
However, their contributions are limited to static processes or stationary time series \citep{follmann1995approximate,ibrahim2009missing,fitzmaurice2000generalized,troxel1998analysis}. 
A specialized model for potentially non-stationary multivariate time series is the state space model (also known as dynamic linear model \citep{aoki2013state}), which has been widely applied in engineering, economics, statistics and many other fields \citep{schmidt1966application,harvey1990forecasting,scharf1991statistical,gannot1998iterative,kiencke2000automotive,manoliu2002energy,aoki2013state,lai2013adaptive} to estimate time-varying parameters in dynamic systems. 
Even though the state space model tolerates missing values in response variable, it does not permit missing values in explanatory variables \citep{cipra1997kalman,petrics2009}: missing data in explanatory variables must either be imputed or discarded as in complete case analysis.
Only a few missing data imputation approaches have been explored for time series using state space model. 
\citet{bashir2018handling} proposed a state-space-model formatted vector autoregressive model to impute missing values in the response variable only 
 using its past values, assuming stationarity and excluding the influence of exposures or covariates.
\citet{naranjo2013extending} proposed a state-space model algorithm for missing values in exogenous predictors, under the assumption that these variables are independent of the response. 
Both methods require i.i.d subjects for inference, and neither method addresses the commonly encountered missing data scenario in mobile device data -- missingness in outcome variable and explanatory variables induced by including lagged outcomes as regressors. 

Comprehensive assessments of existing imputation methods and the development of novel imputation strategies are in urgent need for missing data imputation in potentially non-stationary multivariate time series. 
We consider an N-of-1 study design, where a single subject constitutes the entire study with recurrent interventions or exposures observed for the same subject.
We propose a Monte Carlo EM algorithm for state space models  (MCEM-SSM) in order to handle missing lagged outcomes used as regressors and to estimate quantities of interest in a non-stationary multivariate time series.
Comprehensive comparisons to commonly used strategies for addressing missing data in terms of bias, estimation error, and coverage are conducted in simulations of both stationary and non-stationary time series.

The structure of the paper is as follows. Section 2 introduces notation of multivariate time series and the associated pattern of missing data. 
The state space model used for the BLS application is introduced in Section 3. Section 4 introduces the proposed Monte Carlo EM algorithm (MCEM-SSM) for handling missing data in (potentially non-stationary) time series. Section 5 presents performance evaluation of the proposed method and compares it to other commonly used methods, using simulations of both stationary and non-stationary time series under various missing mechanisms. 
In Section 6, we apply the proposed method to a multi-year smartphone-based observational study of bipolar and schizophrenia patients, evaluating the relationship between phone-based social connectivity and patients' negative mood. Finally, Section 7 concludes with a discussion of key findings and future directions.


\section{Notations and missing data pattern}
Missing data can be prevalent both in actively collected data (e.g., moods or psychiatric symptoms collected from surveys) as they need participants' engagement, and passively collected data (e.g., telecommunication data, accelerometer data).
In particular for the BLS study, the missing rate for actively collected daily survey data 
ranges from $10\%$ to $90\%$.
Without loss of generality, we aim to address this prevalent missing data scenario in mobile device data, where actively collected data (used as outcomes and explanatory variables in the BLS study) contain missing values while other passively collected data (used also as explanatory variables in the BLS study) are fully observed throughout the follow-up period.

We consider a N-of-1 study design where a single subject is followed over time at $t=1,\ldots,T$. At time $t$, we denote the outcome as $Y_t$, the exposure(s) or treatment(s) as $A_t$, and other covariate(s) as $\C_t$. The observation series of $(Y_t,A_t,\C_t)$,  at $t=1,\ldots,T$, forms a multivariate time series. 
We denote the q-lagged value of the outcome as $Y_{t-q}$ 
for $q>0$. 
We denote missingness indicators $M_t$ for the outcome with $M_t=1 $ if $Y_t$ is missing and $M_t=0$ if $Y_t$ is observed. 
Thus, the set of time points where outcomes are missing is given by $\mathbf{T}_{\text{mis}}=\{t: m_t=1,t=1,\ldots,T\}$ with cardinality $|T_{\text{mis}}|$; similarly, the set of time points where outcomes are observed is given by $\mathbf{T}_{\text{obs}}=\{t: m_t=0,t=1,\ldots,T\}$ with cardinality $|T_{\text{obs}}|$. 
Accordingly, we partition outcomes into a set of observed outcomes $\mathbf{Y}_{\text{obs}} = \{Y_t: m_t=0 \text{ or } t \notin \mathbf{T}_{\text{mis}} \}$ and a set of missing outcomes $\mathbf{Y}_{\text{mis}}=\{Y_t: m_t=1 \text{ or } t \in \mathbf{T}_{\text{mis}} \}$. 
For observed time points $\mathbf{T}_{\text{obs}}$, we further partition it into fully observed time points $\mathbf{T}_{\text{obs}}^{(0)}$, where there is no missing data in both response and explanatory variables for analysis, and partial observed time points $\mathbf{T}_{\text{obs}}^{(1)}$, where the response is observed but explanatory variables contain missing data. Summary table of notations with data example is shown in Appendix Section A.

\begin{figure}[h!]
\centering
\includegraphics[width=\linewidth]{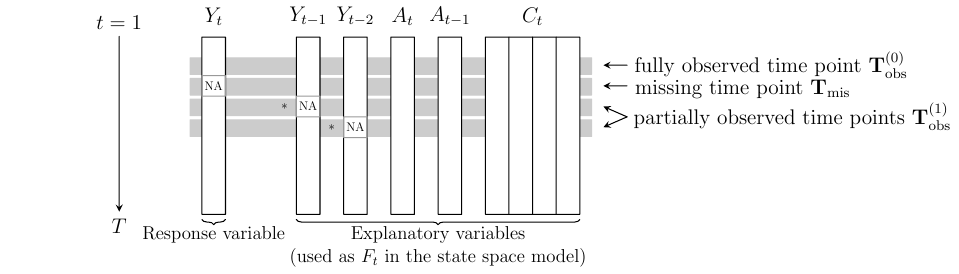}
\caption{Non-monotonic missing data due to lagged outcomes used as explanatory variables for time series analysis. $Y_t$ denotes outcome at time $t$, whereas ($Y_{t-1}$, $Y_{t-2}$) denote two lagged outcomes included as explanatory variables for analysis. $A_t$ and $A_{t-1}$ denote exposures at time $t$ and $t-1$. All other covariates included for analysis are included in $\C_t$. Observations can be partitioned into fully observed time points $\mathbf{T}_{\text{obs}}^{(0)}$, partial observed time points $\mathbf{T}_{\text{obs}}^{(1)}$ (when the response variable is observed but there is missing data in explanatory variables), and missing time points $\mathbf{T}_{\text{mis}}$ (when the response variable is missing). Missing outcome $Y_t$ at time $t$ induces a missing time point at time $t$ and two partially observed time points at $t-1$ and $t-2$ (marked with *) for analysis.}
\label{fig:missing_pattern}
\end{figure}

As lagged outcomes are commonly used as explanatory variables to account for auto-correlation in time series analysis, missingness in the outcome naturally results in missingness in the explanatory variables.  
Figure~\ref{fig:missing_pattern} demonstrated a non-monotone missing data pattern, when lagged outcomes up to 2 lags are included as explanatory variables for analysis. 
In general, the more lagged outcomes are included, the higher the overall missing rate for analysis.

\section{The state space approach}

The state space model has been recognized as one of the most noticeable innovations in engineering \citep{gannot1998iterative,kiencke2000automotive}, econometrics \citep{lai2013adaptive}, mathematical finance \citep{manoliu2002energy}, and time series analysis \citep{harvey1990forecasting,scharf1991statistical}. 
State space modeling provides a flexible framework for time-varying parameters in non-stationary time series with the powerful estimation tools of the Kalman filter and smoothing \citep{kalman1960new,kalman1961new}.
Specifically, the state space model formulates time series $Y_t$, $t=1,2,\ldots$, as observations of a dynamic system's output up to additive Gaussian random noise. 
The ``observational equation'' describes the dependence of this time series on a latent process of hidden states, the evolution of which is described by the ``state equation.'' Denote $\theta_t$ as the $d \times 1$ latent state vector at $t=1,2,\ldots$, and assume it to be a Markov process such that $\theta_t \indep \theta_s | \theta_{t-1}$ for $\{\theta_s: s<t\}$ \citep{kalman1960new,kalman1961new,harvey1990forecasting,scharf1991statistical}.
\begin{defn}[Linear state space model] For $t=1,2,\ldots$, the state equation of linear state space model is
\vspace{-0.1cm}
\begin{equation}
\theta_t = G_t \theta_{t-1} + w_t, \quad w_t \sim N_d(0,Q)\vspace{-0.2cm}
\label{eq:state}	
\end{equation}
where $\theta_{t}$ denotes the $d \times 1$ state vector, $G_t$ is the $d \times d$ state transition matrix, and $w_t$ represents the $d \times 1$ independently and identically distributed noise vector, following distribution  $N_d(0,Q)$. The observational equation of linear state space model is
\vspace{-0.3cm}
\begin{equation}
Y_t=F_t \theta_t + v_t,  \quad v_t \sim N_n(0,R)\vspace{-0.3cm}
\label{eq:obs}
\end{equation}
where $Y_t$ is the $n \times 1$ vector of  outcomes, $F_t$ is the $n \times d $ design matrix, and $v_t$ is the i.i.d observational noise vector, following distribution $N_n(0, R)$.
\end{defn}

\subsection{State space model for the BLS application}

In the Bipolar Longitudinal Study (BLS), the outcome variable is a one-dimensional ($n=1$) summary for self-reported negative mood. Note that the extension to multiple correlated outcomes is not trivial, we start with a one-dimension outcome, compatible with the BLS application and N-of-1 study setting.

\begin{figure}[h!]
\centering
\includegraphics[width=0.8\linewidth]{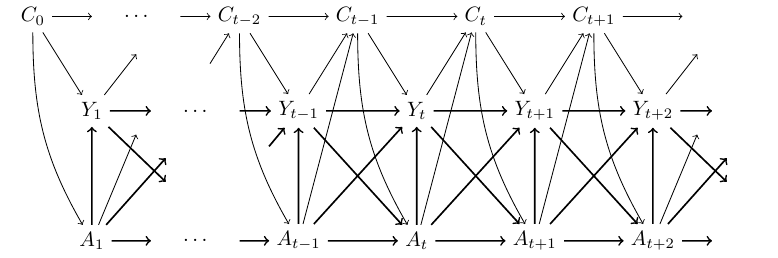}
\caption{Time series directed acyclic graph (DAG) (also known as time series chain graph) displaying the temporal order and relationship of time-varying exposures, outcomes, and covariates. 
Outcome $Y_t$ depends on its previous value $Y_{t-1}$, current and previous exposures $\A_{(t-1):t}$, and previous covariates $\C_{t-1}$; exposure $A_t$ depends on the previous exposure $A_{t-1}$, outcome $Y_{t-1}$, and covariates $\C_{t-1}$; covariate(s) $\C_t$ depend on their previous value(s) $\C_{t-1}$, as well as the most recent exposure $A_{t-1}$ and outcome $Y_{t-1}$.}
\label{fig:dag1}	
\end{figure}

As current mood may depend on past mood, exposure, and other covariates, we consider a linear relationship of outcome $Y_t$ on its previous value $Y_{t-1}$, current and previous exposures $(A_t, A_{t-1})$, and current covariates $C_t$ as illustrated in the causal diagram in Figure~\ref{fig:dag1}. The specified statistical model (also known as data generation process) for $Y_t$ is 
\vspace{-0.3cm}
\begin{equation}
Y_t =  \beta_{0,t} + \rho_t Y_{t-1} + \beta_{1,t} A_t +\beta_{2,t} A_{t-1} + \beta_{c,t} C_t + v_t 
\vspace{-0.3cm}\label{eq:ssmmodel}
\end{equation}
which can be expressed in state space model form as
\vspace{-0.3cm}
\begin{equation}
Y_t = F_t \theta_t +  v_t \vspace{-0.3cm}
\end{equation}
where design matrix $F_t =(1,Y_{t-1},A_t,A_{t-1},C_t)$ contains lagged outcomes and current and past exposures and covariates, and the hidden states $\theta_t=(\beta_{0,t},\rho_t,\beta_{1,t},\beta_{2,t},\beta_{c,t})'$ represents the time-varying unknown regression coefficients at time $t$. A more general case will be introduced in Section 4 with detailed explanation in Appendix Section C.

If $\theta_t=\theta$ is time-invariant and $A_t$ and $C_t$ are stationary time series, the resulting time series $Y_t$ is also stationary. 
On the other hand, $Y_t$ may follow a non-stationary or dynamic process, when certain components of $\theta_t$ are time-varying (e.g., random walk, AR(p) process, or periodic stable) or when the variance of observational noise changes over time. 
Time-varying scenarios are prevalent in psychiatric research, where the mechanisms underlying complex mental health conditions are often not well understood. 
A time-varying intercept may be needed to capture irregular changes in those conditions due to unmeasured factors.
Similarly, treatment effects can fluctuate in response to major life changes (e.g, a change of job or disease severity). 
In such cases, treatment effects can be modeled as periodic-stable processes, which is time-invariant within periods but differs across periods. 
The state space model provides a generic mathematical framework that can represent both stationary (static) and non-stationary (dynamic) time series, and can be easily modified to accommodate more complex dynamics than those illustrated in Figure~\ref{fig:dag1} and Equation~\eqref{eq:ssmmodel}. 


\section{Monte Carlo EM algorithm of the state space model with missing data (MCEM-SSM)}
Likelihood-based methods combined with the EM algorithm have long been employed for parameter estimation with missing data. 
Existing approaches include complete case analysis, traditional imputation techniques such as multiple imputation, and static state-space models designed to impute missing values either in the response variable \citep{petrics2009} or exogenous covariates independent from the outcome \citep{naranjo2013extending}. 
However, mobile device data with intensive longitudinal measurements and complex temporal dependencies pose a distinct challenge: simultaneous missingness in both response and explanatory variables (as shown in Figure~\ref{fig:missing_pattern}). 
We propose a Monte Carlo Expectation-Maximization algorithm within a state-space modeling framework (MCEM-SSM), to address this complex missing data structure in potentially non-stationary time series.

The MCEM-SSM approach iterates between two main steps until convergence: the Expectation (E) step and the Maximization (M) step. 
In the E step, we calculate the expected log-likelihood by integrating over the missing values, conditioned on the observed data. Components without closed-form solutions are approximated using Monte Carlo simulations. 
In the M step, we maximize the expected log-likelihood obtained in the E step to update parameter estimation. 
The optimization process utilizes analytical solutions whenever possible to improve computational efficiency. When closed-form solutions are unavailable, numerical optimization techniques are employed to ensure convergence.
Our proposed algorithm assumes Gaussianity and a linear state space framework \citep{ibrahim2005missing,little2019statistical}. However, Kalman recursions also extend to non-Gaussian distributions and non-linear system, enabling further adaptations of the algorithm to accommodate other distributions.

In a general setting, the outcome $Y_t$  depends on previous outcomes $\Y_{(t-q):(t-1)}=(Y_{t-1},\ldots, Y_{t-q})$ with $q \ge 1$, current and previous exposures $\A_{(t-p):t} = (A_{t},\ldots,A_{t-p})$ with $p \ge 0$, and current and previous covariates $\C_{(t-o):t}=(\C_t,\ldots,\C_{t-o})$ with $o \ge 0$. Equation \eqref{eq:ssmmodel} is a specific case  selected for the BLS application with $q=1, p=1, o=0$ \citep{valeri2023digitalpsychiatry,fowler2022,cai2024causal}. A detailed explanation for the general setting is shown in Appendix Section C.

\subsection{MCEM-SSM algorithm}

For state space models, the likelihood of the complete data when all observations and hidden states $\boldsymbol{\theta}_{1:T}=(\theta_1,\ldots,\theta_T)$ are observed on ignorable missing mechanism is
\begin{equation}
\begin{split}
f_{\Theta}(\mathbf{Y}_{obs},\mathbf{A}_{1:T},\mathbf{C}_{1:T},\boldsymbol{\theta}_{1:T}) & = f_{\mu_0,\Sigma_0}(\theta_0) \prod_{t=1}^T f_{Q}(\theta_t|\theta_{t-1}) \prod_{t \in \mathbf{T}_{obs}} f_{R}(y_t|\theta_t)
\vspace{-0.3cm}	
\end{split}
\label{eq:likelihood}
\end{equation}
with $\Theta = (\mu_0,\Sigma_0,Q,R)$. Assuming normality and ignoring the constant terms, the natural logarithm of the complete data likelihood in the above equation \eqref{eq:likelihood} can be written as
\vspace{-0.3cm}
\begin{equation}
\begin{split}
& lnL(\Theta|\mathbf{Y}_{obs},\mathbf{A}_{1:T},\mathbf{C}_{1:T},\boldsymbol{\theta}_{1:T})  = \frac{1}{2} ln (\text{det}(\Sigma_0^{-1})) - \frac{1}{2} (\theta_0 -\mu_0)' \Sigma_0^{-1} (\theta_0 -\mu_0) \\
& \quad\quad\quad\quad\quad\quad\quad\quad + \frac{1}{2} \sum_{t=1}^T ln (\text{det}(Q^{-1})) - \frac{1}{2}\sum_{t=1}^T (\theta_t - G_t \theta_{t-1})' Q^{-1} (\theta_t - G_t \theta_{t-1}) \\ 
& \quad\quad\quad\quad\quad\quad\quad\quad + \frac{1}{2} \sum_{t \in \textbf{T}_{obs}} ln( \text{det}(R^{-1})) - \frac{1}{2} \sum_{t=1}^T (y_t - F_t \theta_{t})' R^{-1}  (y_t - F_t \theta_{t})
\vspace{-0.3cm}
\end{split}
\label{eq:loglikelihood}
\end{equation}
Note that derivations of \eqref{eq:likelihood} and \eqref{eq:loglikelihood} assume on ignorable missing mechanism and thus are applicable to MCAR (missing completely at random) and MAR (missing at random), with the potential to be extended to MNAR (missing not random) if modeling for the missing mechanism is also included in the likelihood. A detailed discussion is presented in Appendix section B.

To simplify the subsequent derivation, we denote the observed part as $\bO_{obs} = \{\mathbf{Y}_{obs}, \A_{1:T},\C_{1:T} \}$ and the missing part as $\{\mathbf{Y}_{mis}, \boldsymbol{\theta}_{1:T}\}$. 
We partition the observed outcomes $\mathbf{Y}_{obs}$ based on fully and partially observed time points, $\mathbf{Y}_{\text{obs}}^{(0)} = \{Y_t: t \in \mathbf{T}_{\text{obs}}^{(0)}\}$ and $\mathbf{Y}_{\text{obs}}^{(1)} = \{Y_t: t \in \mathbf{T}_{\text{obs}}^{(1)}\}$, as illustrated in Figure~\ref{fig:missing_pattern}.
Summary table of notations with data example is shown in Appendix Section A. 
For partially observed time points $t \in \mathbf{T}_{\text{obs}}^{(1)}$, we partition the hidden states $\theta_t$ into two components: the part corresponding to the observed explanatory variables, $\theta_t^{(0)}$, and and the part corresponding to the missing explanatory variables, $\theta_t^{(1)}$; similarly, we partition the design matrix $F_t$ into two components: the part of observed explanatory variables $F_t^{(0)}$ and the part of missing explanatory variables $F_t^{(1)}$.
For example, in the model \eqref{eq:ssmmodel} specified for the BLS, $y_{t-1}$ is a missing explanatory variable, and thus $F_t^{(1)}=y_{t-1}$ and its corresponding coefficient $\rho_{t}$ is included in $\theta_t^{(1)}=\rho_{t}$; while exposures and covariates are observed, and thus $F_t^{(0)} = (1, A_t, A_{t-1}, \C_t)'$ and their corresponding coefficients are contained in $\theta_t^{(0)}=(\beta_{0,t},\beta_{1,t},\beta_{2,t},\beta_{c,t})'$. 
Explanation for the general case is also provided in Appendix section C.

The EM algorithm begins with the E-step, calculating the expected log-likelihood regarding the conditional distribution of the missing data giving observed data and the current estimates of the unknown parameters from the previous iteration, $\{\Y_{mis}^{(0)},\boldsymbol{\theta}_{1:T}|\bO_{obs}\}$. 
At each iteration $j$, the E-step computes the expected log-likelihood as:
\vspace{-0.3cm}
\begin{equation}
\mathcal{Q}\left(\Theta \mid \Theta^{(j-1)}\right) = \mathbb{E}\left[\ln L(\Theta \mid \mathbf{Y}_{obs}, \mathbf{Y}_{mis}^{(1)}, \A_{1:T},\C_{1:T}, \boldsymbol{\theta}_{1:T}) \mid \mathbf{Y}_{obs}, \A_{1:T},\C_{1:T}, \Theta^{(j-1)}\right],\vspace{-0.3cm}
\end{equation}
where $\Theta^{(j-1)} = (\mu_0^{(j-1)},\Sigma_0^{(j-1)},Q^{(j-1)},R^{(j-1)})$ are the estimated parameters at the previous iteration $j-1$.
Plugging in the log-likelihood specified in Equation (5), we have
\vspace{-0.1cm}
\begin{footnotesize}
\begin{equation}
\begin{split}
& \mathcal{Q}\left(\Theta \mid \Theta^{(j-1)}\right) = \mathbb{E}\left[\ln L(\Theta \mid \mathbf{Y}_{obs}, \mathbf{Y}_{mis}^{(1)}, \A_{1:T},\C_{1:T}, \boldsymbol{\theta}_{1:T}) \mid \mathbf{Y}_{obs}, \A_{1:T},\C_{1:T}, \Theta^{(j-1)}\right] \\
& =\frac{1}{2} \ln ( \text{det}(\Sigma_0^{-1})) +  \frac{T}{2} \ln ( \text{det}(Q^{-1}))   + \frac{|\mathbf{T}_{obs}|}{2} \ln ( \text{det}( R^{-1})) \\
& -\frac{1}{2} \operatorname{tr} \left\{ \Sigma_0^{-1} \left( \mathbf{P}_0^T+ (\theta_0^T-\mu_0) (\theta_0^T-\mu_0)' \right) \right\}\\
& - \frac{1}{2}  \operatorname{tr}\left\{ Q^{-1} \left[ \sum_{t=1}^T (\mathbf{P}_{t}^T + \theta_t^T \theta_t^{'T})- \sum_{t=1}^T (\mathbf{P}_{t,t-1}^T + \theta_t^T \theta_{t-1}^{'T})\Phi' - \sum_{t=1}^T \Phi (\mathbf{P}_{t-1,t}^T +  \theta_{t-1}^T \theta_{t}^{'T}) + \sum_{t=1}^T \Phi (\mathbf{P}_{t-1}^T+\theta_{t-1}^T\theta_{t-1}^{'T}) \Phi' \right] \right\} \\
& - \frac{1}{2} \operatorname{tr} \left\{  \sum_{t\in T_{obs}^{(0)}} R^{-1} (y_t-F_t \theta_t^T)(y_t-F_t \theta_t^T)' +  \sum_{t\in T_{obs}^{(0)}} R^{-1} F_t \mathbf{P}_{t}^T F'_t  \right\}\\
& - \frac{1}{2} \operatorname{tr} \left\{ \sum_{t\in T_{obs}^{(1)}} R^{-1}  (y_t-F_t^{(0)}\theta_t^{T(0)}- \E[ F_t^{(1)} \theta_t^{(1)} \mid \bO_{obs} ]) (y_t-F_t^{(0)} \theta_t^{T(0)}- \E[ F_t^{(1)} \theta_t^{(1)} \mid \bO_{obs} ])' \right\} \\
& - \frac{1}{2} \operatorname{tr} \left\{ \sum_{t\in T_{obs}^{(1)}} R^{-1} F_t^{(0)}\mathbf{P}_{t}^{T(0)}F_t^{'(0)} \right\} \\
& - \frac{1}{2} \operatorname{tr} \left\{ \sum_{t\in T_{obs}^{(1)}}  R^{-1} \E \left[ F_t^{(0)}(\theta_t^{(0)} - \theta_t^{T(0)} ) (F_t^{(1)} \theta_t^{(1)}- \E[ F_t^{(1)} \theta_t^{(1)} \mid \bO_{obs} ])' \mid  \bO_{obs} \right]\right\} \\
& - \frac{1}{2} \operatorname{tr} \left\{ \sum_{t\in T_{obs}^{(1)}} R^{-1} \E \left[ (F_t^{(1)} \theta_t^{(1)}- \E[ F_t^{(1)} \theta_t^{(1)} \mid \bO_{obs} ]) (\theta_t^{(0)} - \theta_t^{T(0)} )'F_t^{'(0)}  \mid  \bO_{obs} \right]\right\} \\
& - \frac{1}{2}  \operatorname{tr} \left\{ \sum_{t\in T_{obs}^{(1)}}  R^{-1} \E \left[ (F_t^{(1)} \theta_t^{(1)}- \E[ F_t^{(1)} \theta_t^{(1)} \mid \bO_{obs} ]) (F_t^{(1)} \theta_t^{(1)}- \E[ F_t^{(1)} \theta_t^{(1)} \mid \bO_{obs} ])' \mid  \bO_{obs} \right]\right\} ,
\end{split}
\label{eq:Q_likelihood}
\end{equation}
\end{footnotesize}
where $\theta_t^T = \E[\theta_t| \bO_{obs}]$, $\theta_t^{T(0)} = \E[\theta_t^{(0)}| \bO_{obs}]$,
$\mathbf{P}_{t_1,t_2}^T = \E[(\theta_{t_1} - \theta_{t_1}^T )(\theta_{t_2} - \theta_{t_2}^T)'| \bO_{obs}]$ and $\mathbf{P}_{t_1,t_2}^{T(0)} = \E[(\theta_{t_1}^{(0)} - \theta_{t_1}^{T(0)} )(\theta_{t_2}^{(0)} - \theta_{t_2}^{T(0)})'| \bO_{obs}]$ with simplified notations $\mathbf{P}_{t}^T$ and $\mathbf{P}_{t}^{T(0)}$ when $t_1=t_2=t$. Note that expectations in \eqref{eq:Q_likelihood} do not have closed-form solutions and are approximated using Monte Carlo simulation via MCMC algorithm as follows 
\vspace{-0.1cm}
\begin{footnotesize}
\[ \tilde{\theta}_0^T = \frac{1}{M} \sum_{m=1}^M \theta_0^{[m,j]},\quad \tilde{\theta}_t^T = \frac{1}{M} \sum_{m=1}^M \theta_t^{[m,j]},\quad \tilde{\theta}_t^{T(0)} = \frac{1}{M} \sum_{m=1}^M \theta_t^{(0)[m,j]}\]
\vspace{-0.5cm}
\[ \tilde{\mathbf{P}}_{t_1,t_2}^T = \frac{1}{M}\sum_{m=1}^M (\theta_{t_1}^{[m,j]} - \tilde{\theta}_{t_1}^T)(\theta_{t_2}^{[m,j]} - \tilde{\theta}_{t_2}^T)', \quad \tilde{\mathbf{P}}_{t_1,t_2}^{T(0)} = \frac{1}{M}\sum_{m=1}^M (\theta_{t_1}^{(0)[m,j]} - \tilde{\theta}_{t_1}^{T(0)})(\theta_{t_2}^{(0)[m,j]} - \tilde{\theta}_{t_2}^{T(0)})'\]
\[ \tilde{\E}[ F_t^{(1)} \theta_t^{(1)} \mid \bO_{obs} ] =  \frac{1}{M} \sum_{m=1}^M  F_t^{(1),[m,j]}\theta_t^{(1),[m,j]}\] 
\begin{equation*}
	\begin{split}
	& \tilde{\E} \left[F_t^{(0)} (\theta_t^{(0)} - \theta_t^{T(0)} ) (F_t^{(1)} \theta_t^{(1)}- \E[ F_t^{(1)} \theta_t^{(1)} \mid \bO_{obs} ])' \mid  \bO_{obs} \right]\\
	& = \frac{1}{M} \sum_{m=1}^M  F_t^{(0)} \left( \theta_t^{(0),[m,j]} - \frac{1}{M} \sum_{m=1}^M \theta_t^{(0),[m,j]}  \right) \left(F_t^{(1),[m,j]}\theta_t^{(1),[m,j]}-\frac{1}{M} \sum_{m=1}^M F_t^{(1),[m,j]}\theta_t^{(1),[m,j]} \right)'
	\end{split}
	\end{equation*}
	\begin{equation*}
	\begin{split}	
& \tilde{\E} \left[ (F_t^{(1)} \theta_t^{(1)}-  \E[ F_t^{(1)} \theta_t^{(1)} \mid \bO_{obs} ] ) (\theta_t^{(0)} - \theta_t^{T(0)})' F_t^{'(0)} \mid  \bO_{obs} \right] \\
& = \frac{1}{M} \sum_{m=1}^M \left(F_t^{(1),[m,j]}\theta_t^{(1),[m,j]}-\frac{1}{M} \sum_{m=1}^M F_t^{(1),[m,j]}\theta_t^{(1),[m,j]} \right)  \left( \theta_t^{(0),[m,j]} - \frac{1}{M} \sum_{m=1}^M \theta_t^{(0),[m,j]}  \right)' F_t^{'(0)} 
	\end{split}
	\end{equation*}
	\begin{equation*}
	\begin{split}
	& \tilde{\E} \left[ (F_t^{(1)} \theta_t^{(1)}- \E[ F_t^{(1)} \theta_t^{(1)} \mid \bO_{obs} ]) (F_t^{(1)} \theta_t^{(1)}- \E[ F_t^{(1)} \theta_t^{(1)} \mid \bO_{obs} ])' \mid  \bO_{obs} \right] \\
	& = \frac{1}{M} \sum_{m=1}^M \left(F_t^{(1),[m,j]}\theta_t^{(1),[m,j]}-\frac{1}{M} \sum_{m=1}^M F_t^{(1),[m,j]}\theta_t^{(1),[m,j]} \right)\left(F_t^{(1),[m,j]}\theta_t^{(1),[m,j]}-\frac{1}{M} \sum_{m=1}^M F_t^{(1),[m,j]}\theta_t^{(1),[m,j]} \right)'
	\end{split}
	\end{equation*}
\end{footnotesize}
where $\boldsymbol{\theta}_{1:T}^{[m,j]}$ and $\mathbf{Y}_{mis}^{[m,j]}$ used in the design matrix $F_t^{(1),[m,j]}$ for $m=1,2,\ldots, M$ are MCMC dependent samples based on $\Theta^{(j)} = (\mu_0^{(j)},\Sigma_0^{(j)},Q^{(j)},R^{(j)})$. It is common to use smaller values of $M$ to start and increase it as the estimation gets closer to the MLE in MCEM algorithm \cite{booth1999maximizing}. 

The maximization (M) step maximizes the derived expected log-likelihood in Equation \eqref{eq:Q_likelihood}, at each iteration $j$ and estimate unknown parameter $\Theta$ with closed solution \citep{anderson1958introduction, naranjo2013extending}. Setting the derivatives with respect to each component of  $\Theta^{(j-1)} = (\mu_0^{(j-1)},\Sigma_0^{(j-1)},R^{(j-1)},Q^{(j-1)})$ to 0, we are able to solve the updated parameter estimation for the next iteration as follows:
\begin{footnotesize}
\begin{equation*}
\begin{split}
\mu_0^{(j)} & =\tilde{\theta}_0^T \\
\Sigma_0^{(j)} & = \mathbf{P}_{0}^{T}+\left(\tilde{\theta}_{0}^{T}-\mu_0\right)\left(\tilde{\theta}_{0}^{T}-\mu_0 \right)' \\
Q^{(j)} & = \frac{1}{T}\left[ \sum_{t=1}^T (\mathbf{P}_{t}^T + \tilde{\theta}_t^T \tilde{\theta}_t^{'T})- \sum_{t=1}^T (\mathbf{P}_{t,t-1}^T + \tilde{\theta}_t^T \tilde{\theta}_{t-1}^{'T})\Phi' - \sum_{t=1}^T \Phi (\mathbf{P}_{t-1,t}^T +  \tilde{\theta}_{t-1}^T \tilde{\theta}_{t}^{'T}) + \sum_{t=1}^T \Phi (\mathbf{P}_{t-1}^T+\tilde{\theta}_{t-1}^T\tilde{\theta}_{t-1}^{'T}) \Phi' \right]\\
R^{(j)} & = \frac{1}{T_{obs}} \Biggr[  \sum_{t\in T_{obs}^{(0)}}  (y_t-F_t \tilde{\theta}_{t}^T)(y_t-F_t \tilde{\theta}_{t}^T)'  +  \sum_{t\in T_{obs}^{(0)}} F_t \tilde{\mathbf{P}}_{t}^T F'_t + \sum_{t\in T_{obs}^{(1)}}F_t^{(0)}\tilde{\mathbf{P}}_{t}^{T(0)}F_t^{'(0)} \\ 
& + \sum_{t\in T_{obs}^{(1)}}  (y_t-F_t^{(0)}\tilde{\theta}_t^{T(0)}- \E[ \tilde{F}_t^{(1)} \tilde{\theta}_t^{(1)} \mid \bO_{obs} ]) (y_t-F_t^{(0)} \tilde{\theta}_t^{T(0)}- \E[ \tilde{F}_t^{(1)} \tilde{\theta}_t^{(1)} \mid \bO_{obs} ])' \\
& + \sum_{t\in T_{obs}^{(1)}} \tilde{\E} \left[F_t^{(0)} (\theta_t^{(0)} - \theta_t^{T(0)} ) (F_t^{(1)} \theta_t^{(1)}- \E[ F_t^{(1)} \theta_t^{(1)} \mid \bO_{obs} ])' \mid  \bO_{obs} \right] \\
& +  \sum_{t\in T_{obs}^{(1)}} \tilde{\E} \left[ (F_t^{(1)} \theta_t^{(1)}- \E[ F_t^{(1)}\theta_t^{(1)} \mid \bO_{obs} ] ) (\theta_t^{(0)} - \theta_t^{T(0)})' F_t^{'(0)} \mid  \bO_{obs} \right] \\
& +  \sum_{t\in T_{obs}^{(1)}}   \tilde{\E} \left[ (F_t^{(1)} \theta_t^{(1)}- \E[ F_t^{(1)} \theta_t^{(1)} \mid \bO_{obs} ]) (F_t^{(1)} \theta_t^{(1)}- \E[ F_t^{(1)} \theta_t^{(1)} \mid \bO_{obs} ])' \mid  \bO_{obs} \right]  \Biggr]
\end{split}
\end{equation*}
\end{footnotesize}
The details of the derivation and MCMC approximation are shown in Appendix Section D. Below, we summarize the computational steps of the MCEM-SSM algorithm. 
Inference for the hidden states and unknown parameters is conducted by calculating the $95\%$ credible intervals from the MCMC chain. Note that the algorithm may not necessarily increase the observed data likelihood at each iteration, as the $\tilde{\mathcal{Q}}\left(\Theta \mid \Theta^{(j-1)}\right)$ function is approximated. 

\vspace{0.3cm}
\fbox{
\begin{minipage}{0.9\textwidth}
\paragraph{MCEM-SSM Algorithm} For each iteration $j$, repeat the following E step as:
\begin{small}
\begin{enumerate}
	\item \textbf{Initialize:} Set $m=0$ and initialize missing lagged outcome $\mathbf{Y}_{mis}^{[m,j]}$ (used in the design matrix $F_t^{[m,j]}$) and increase $m$ by 1.
	\item \textbf{Sample hidden states:} For each $m$, sample hidden states $\theta_t^{[m,j]}$ from the conditional distribution $f(\theta_t| \bO_{obs},F_t^{(1),[m-1,j]})$ for $t=0,1,\ldots,T$, using forward filtering backward sampling (BFBS). 
	\item \textbf{Sample missing outcome:} Sample missing lagged outcomes $\mathbf{Y}_{mis}^{[m,j]}$ (used in $F_t^{(1),[m,j]}$) from the conditional distribution $f(\mathbf{Y}_{mis}| \bO_{obs},\boldsymbol{\theta}_{0:T}^{[m,j]})$ using Kalman predictions, and increase $m$ by 1.
	\item \textbf{Iterate:} Repeat steps 2 and 3 until $m=M$ to obtain MCMC dependent samples of missing outcomes and hidden states, $\{\mathbf{Y}_{mis},\boldsymbol{\theta}_{1:T}|\bO_{obs}\}$, to approximate expectations in $\mathcal{Q}\left(\Theta \mid \Theta^{(j-1)}\right)$. 
\end{enumerate}
\end{small}
and M step as:
\begin{small}
\begin{enumerate}
\item \textbf{Maximize:} Update parameter estimation by Maximizing derived $\tilde{\mathcal{Q}}\left(\Theta \mid \Theta^{(j-1)}\right)$ at iteration j.
\end{enumerate}
\end{small}
until convergence.
\end{minipage}
}
\vspace{0.3cm}

Note that the proposed algorithm is applicable to both stationary and non-stationary time series. For stationary data generation process, the MCEM algorithm yields optimal parameter estimates, minimizing mean squared error. For dynamic data generation process and non-stationary multivariate time series, the MCEM algorithm provides proper parameter estimation when coefficients may change over time (e.g,  random walk, AR(p) process, and periodic-stable process). 
Critical change points for periodic-stable parameters are identified simultaneously with popular change point detection algorithm \citep{changepoint} with details shown in Appendix Section F.3.
This capability of identifying change points is particularly valuable in psychiatric research to understand how exposure effects evolve over time -- such as in response to medication changes or major life events.

Additionally, the algorithm imposes no constraints on the structure of the used state space model: parameters can be modeled as a random walk, an AR(p) process, periodic-stable or time-invariant. In practice, the structure of state space model, including unknown parameters structuring time-varying components (e.g, changepoints of period-stable process, or random walk process) is estimated and gradually revealed during the iterative MCEM-SSM process and stabilizes upon convergence. Convergence is achieved when likelihood, estimated coefficients, and other structural parameters describing the time-varying behaviors of the system differ by less than a pre-specified criteria between iterations. Details on convergence and results with different starting points are provided in Appendix Section F.4. 

\section{Simulation}
\label{sec:simulation}

\subsection{Simulation setup}
We evaluate the performance of the proposed MCEM-SSM algorithm using simulations of both stationary and non-stationary multivariate time series of 1000 time points, emulating the typical 2-4 years follow-up period in the Bipolar Longitudinal Study (BLS). 
Following the temporal causal relationship assumed in Figure~\ref{fig:dag1}, we consider that the outcome $Y_t$ is auto-correlated with its previous value $Y_{t-1}$ and depends on exposures on the same day (to represent a contemporaneous or short-term association) and the previous day (to represent a lagged or long-term association), in the presence of other confounders in $C_t$, as shown in \eqref{eq:dgp}. \vspace{-0.3cm}
\begin{equation}
	Y_t = \beta_{0,t} + \rho_t Y_{t-1} + \beta_{1,t} A_t +\beta_{2,t} A_{t-1} + \beta_{c,t} C_t + v_t \text{, } \quad v_t \sim N(0,R) \vspace{-0.3cm}
	\label{eq:dgp}
\end{equation}
We consider two general scenarios: 1) the stationary scenario, in which all covariate coefficients are fixed over time as $\beta_{0,t}=40$, $\rho_t=0.5$, $\beta_{1,t}=-1.5$, $\beta_{2,t}=-0.5$, $\beta_{c,t}=-1$, and $R=0.1$, so that the resulting outcome time series is stationary when $A_t$ and $C_t$ are stationary, and 2) a non-stationary scenario in which certain coefficients are time-varying, including (i) a random walk intercept $\beta_{0,t} = 40 + \beta_{0,t-1} + w_{t}$, $w_t \sim N(0,1)$, (ii) a periodic stable coefficient $\beta_{1,t}$ as \vspace{-0.3cm}
\begin{equation}
	\beta_{1,t}= \begin{cases} 
      -1, & t \leq 400 \\
      -2, & 400 < t \leq 700\\
      -1, & x > 700
   \end{cases} \vspace{-0.3cm}
\label{eq:3periodsA}
\end{equation} 
and (iii) other time-invariant coefficients $\rho_t=0.5$, $\beta_{2,t}=-0.5$, $\beta_{c,t}=-1$ and $R=0.1$, so that the resulting outcome time series is non-stationary. 
Note that time-varying structure and magnitude of coefficients are chosen to resemble those revealed in the BLS data application. We consider missing data generated under MCAR, in which there is no relationship between the missingness of the data and any values (observed or missing); under MAR, when the probability of being missing depends on observed data $A_t$ and $C_t$; and MNAR, when the probability of being missing depends on unobserved data $Y_t$.

For each scenario, 500 simulations are conducted to evaluate the performance of the proposed ``MCEM-SSM'' algorithms in coefficient estimation and further compare it to other widely used missing data strategies.
These strategies include complete case analysis (``cc''), mean imputation (``mean''), last-observation-carried-forward imputation (``locf''), linear (``linear'') and spline (``spline'') interpolation, multiple imputation using (``mp''), and imputation using best-fitted ARIMA models (``arima'').
The Multiple Imputation by Chained Equations (MICE) is implemented for the multiple imputation strategy (``mp'') with details in Appendix Section E. 
Missing data are imputed for both the response variable and lagged outcomes used as explanatory variables. 
For the stationary case, coefficient estimation is performed using a linear regression model (correct model) after imputation. 
For the non-stationary case, we first estimate coefficients using linear regression (an incorrect model for non-stationary data) and then apply the state space model (the correct model).
We investigate imputation strategy performances under various missing mechanisms of MCAR, MAR, and MNAR and under varying missing rates of $25\%$, $50\%$, and $75\%$.

\subsection{Simulation result and comparisons to other missing data methods}

In the following section, we focus on illustrating the estimation of $\beta_{2,t}$ (the coefficient of $A_{t-1}$) as it exhibits the most pronounced bias when comparing the stationary and non-stationary cases. 
Further explanation is provided in the subsequent paragraphs. 
Similar estimation comparison plots for other coefficients ($\beta_{0,t}$, $\beta_{1,t}$, and $\beta_{c,t}$) under MCAR, MAR, and MNAR and under different missing rate are shown in Appendix Section F.

\begin{figure}[ht]
    \centering 
    \includegraphics[width=\linewidth]{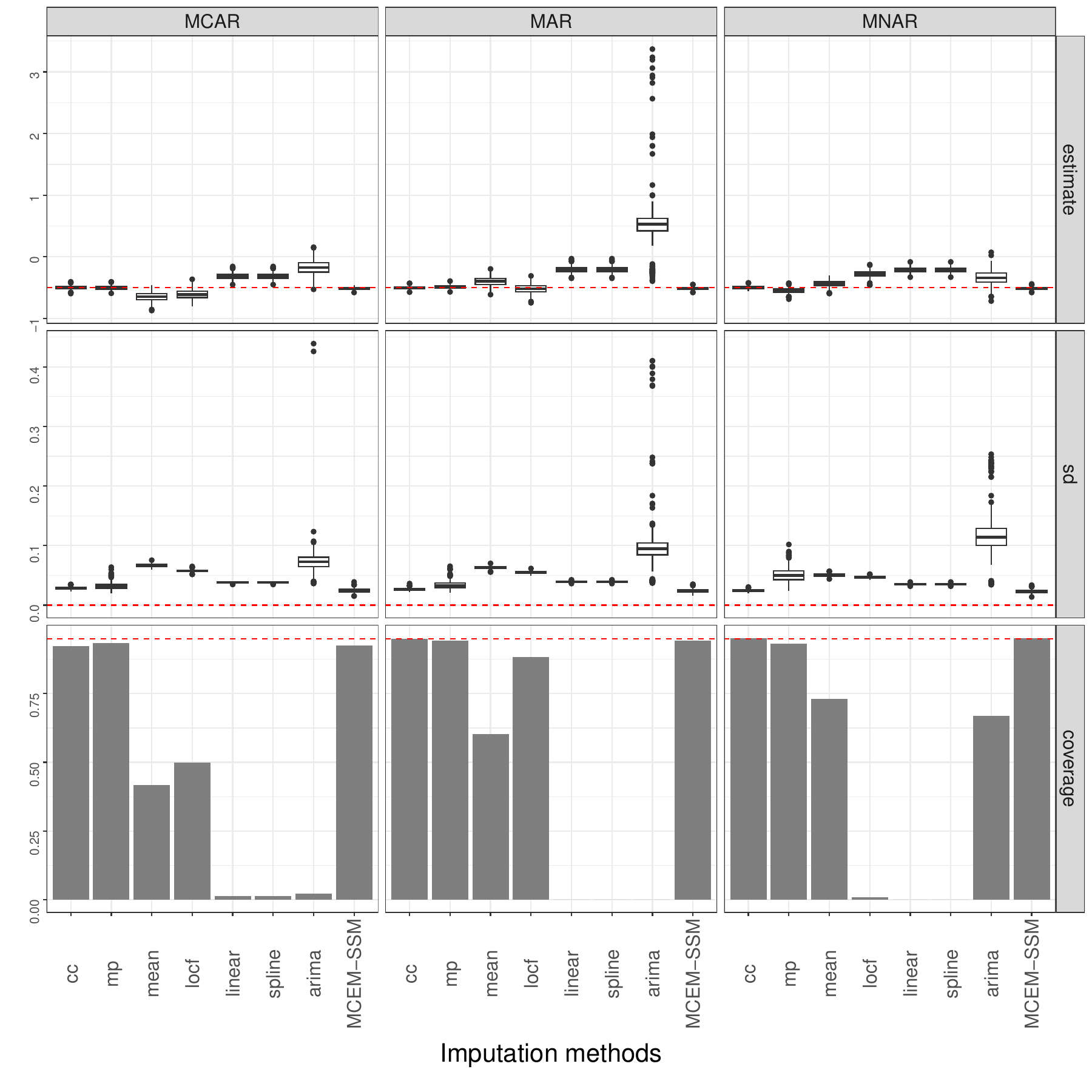}
\caption{Boxplots of the estimate (top), standard error (middle), and $95\%$ CI coverage (bottom) of the coefficient of $A_{t-1}$ for the stationary scenario, over 500 simulations under MCAR (left), MAR (middle), and MNAR (right) and missing rate of $50\%$.
Methods include complete case analysis (``cc''), multiple imputation (``mp''), mean imputation (``mean''), last-observation-carried-forward (``locf''), linear interpolation (``linear''), spline interpolation (``spline''), imputation with best ARIMA model (``arima'') under linear analytical models 
as well as the proposed ``MCEM-SSM'' algorithm.}
\label{fig:simulation_x_1}
\end{figure}

For the stationary case, simulation results align with literature on the performance of widely-used imputation approaches \citep{spratt2010strategies,white2010bias}. 
Figure~\ref{fig:simulation_x_1} illustrates the estimation result for $\beta_{2,t}$ (coefficient of $A_{t-1}$) in terms of bias, standard error, and coverage of $95\%$ confidence interval (CI), using various missing data imputation strategies, under MCAR, MAR, and MNAR with a missing rate of $50\%$. 
Estimation results of other coefficients and missing rate are shown in Appendix Section F, Figures 1-4. For MCAR and MAR,
results confirm that complete case analysis and multiple imputation with correct model specification provide satisfactory coefficient estimation as well as adequate coverage for $95\%$ CI. 
The MCEM-SSM also achieves comparable estimation accuracy and CI coverage.
In contrast, all other imputation strategies demonstrate substantial bias in the coefficient estimation and significant under-coverage for CI, providing clear warnings against their application to intensive longitudinal data or entangled multivariate time series data even in stationary settings. 

\begin{figure}
    \centering 
    \includegraphics[width=\linewidth]{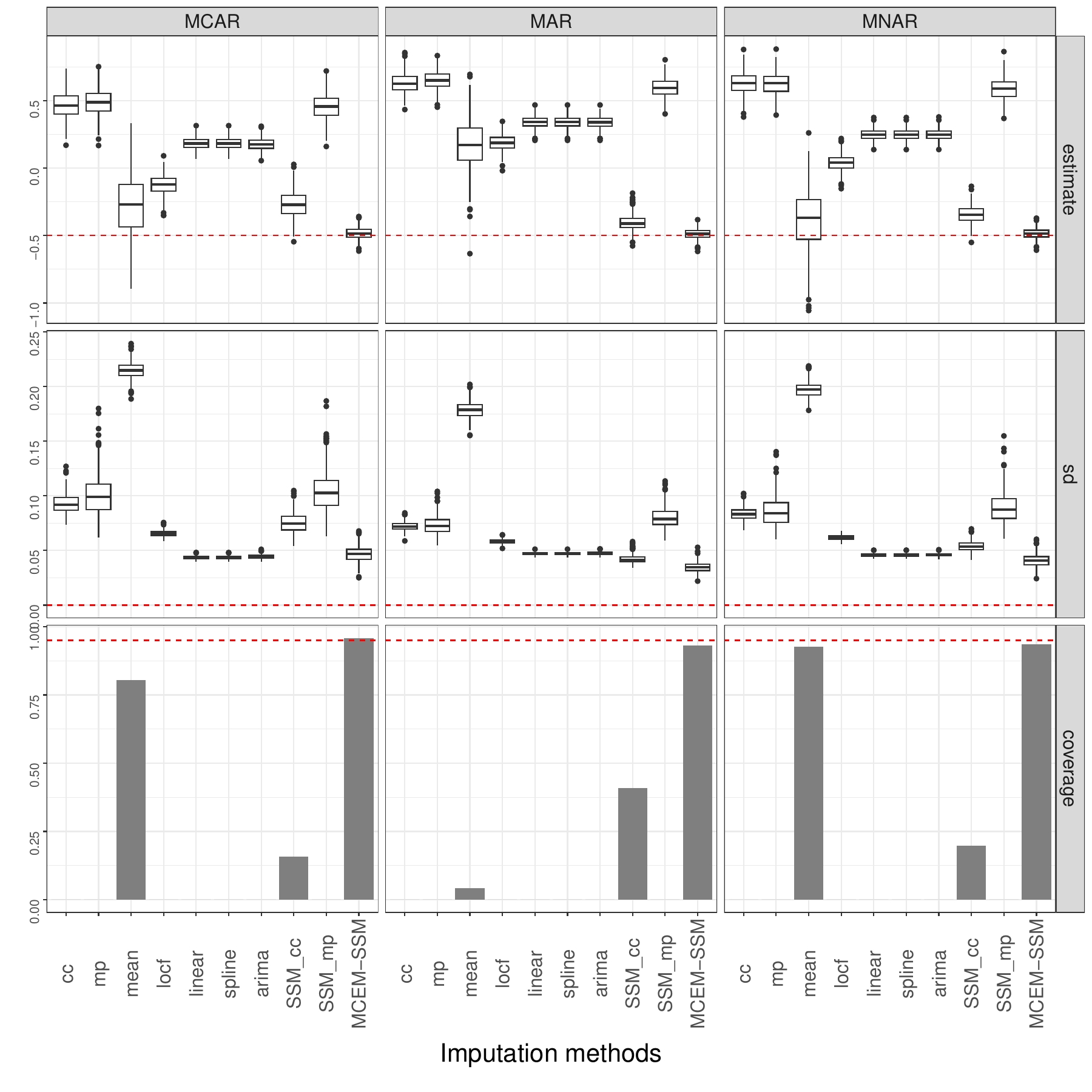}
\caption{Boxplots of the estimate (top), standard error (middle), and $95\%$ CI coverage (bottom) of the time-invariant coefficient of $A_{t-1}$ for the non-stationary scenario, over 500 simulations under MCAR (left), MAR (middle), and MNAR (right) and missing rate of $50\%$.
Methods include complete case analysis (``cc''), multiple imputation (``mp''), mean imputation (``mean''), last-observation-carried-forward (``locf''), linear interpolation (``linear''), spline interpolation (``spline''), imputation with best ARIMA model (``arima'') under linear analytical models, and complete case analysis and multiple imputation  under state space models (``SSM\_cc'' and ``SSM\_mp''), as well as the proposed ``MCEM-SSM'' algorithm.}
\label{fig:simulation_x_1_nonsta}
\end{figure}

For the non-stationary case, Figure~\ref{fig:simulation_x_1_nonsta} shows estimation result for the same time-invariant $\beta_{2,t}$ (coefficient of $A_{t-1}$) in terms of bias, standard error, and coverage of $95\%$ confidence interval under MCAR, MAR, and MNAR with a missing rate of $50\%$. 
Additional results for the remaining time-varying coefficients (random walk intercept $\beta_{0,t}$ and periodic stable $\beta_{1,t}$) and time-invariant coefficients ($\rho_t$ and $\beta_{c,t}$) under MCAR, MAR, and MNAR and under various missing rate of $25\%$, $50\%$, and $75\%$ are shown in Appendix Section F, Figures 5-11. 
Under MCAR and MAR, results confirm that the proposed MCEM-SSM algorithm, when applied with a correctly specified state space model, provides satisfactory estimation for both time-varying and time-invariant coefficients with adequate $95\%$ CI coverage. 
In contrast, all other imputation methods yield biased estimates, even when the correctly specified state space model is used for post-imputation analysis.
To be specific, complete case analysis, which retains only fully observed time points, distorts the original temporal relationships among variables.
As a result, it introduces mild to moderate bias in coefficients related to variables measured at the same time as the outcome (e.g., $\beta_{1,t}$ for $A_t$ and $\beta_{c,t}$ for $C_t$) and induces substantial bias in coefficients of lagged variables (e.g., $\beta_{2,t}$ for $A_{t-1}$ and $\rho_t$ for $Y_{t-1}$).
This finding aligns with those from \citet{carroll1985comparison,gleser1987limiting} -- replacing $Y_{t-1}$ with its autocorrelated surrogate $Y_{t-2}$ leads to biased estimates in coefficients of predictors correlated with $Y_{t-1}$ (i.e. $A_{t-1}$ according to \eqref{eq:dgp}), while leaving uncorrelated predictors (i.e., $A_t$ and $C_t$ according to \eqref{eq:dgp}) relatively unaffected.
Multiple imputation, commonly used for intensive longitudinal data \citep{twisk2002attrition,spratt2010strategies}, also results in significant bias and substantial anti-conservative  CI.
At last, all other widely used imputation methods introduce substantial bias and fail to provide adequate CI coverage. 
In addition, change points for periodic stable coefficients $\beta_{1,t}$ (coefficient of $A_t$) are identified. 
Results of change point detection across different imputation strategies is shown in Appendix Section F, Figure 12. 
All strategies with correctly specified state space model successfully identify change points close to their true values. 
However, complete case analysis performs worse than other approaches, likely due to its reliance on a reduced dataset and the occurrence of missing data near change points.

Note that the satisfactory performance of the MCEM-SSM method under MNAR is attributable to the linearity of the data-generating process specified in Equation~\eqref{eq_dqp}. This result does not generalize to more complex or misspecified models, where MNAR mechanism leads to biased estimation unless the missingness process is correctly modeled.

\section{Application}

\label{sec:application}
The Bipolar Longitudinal Study (BLS) is a mobile health (mHealth) cohort study that has recruited 74 patients with schizophrenia or bipolar illness from the Psychotic Disorders Division at McLean Hospital since February 2016. 
Once recruited, each participant underwent a comprehensive Diagnostic and Statistical Manual of Mental Disorders (DSM-IV) examination \citep{american2013diagnostic}. 
The study aimed at following each participant for at least 1 year (duration of mean(sd) 357.8(97.9) in days). A rich collection of data about physical activity, GPS locations, and anonymized basic information of calls and texts was passively collected using smartphones via the ``Beiwe'' platform \cite{huang2019activity,barnett2020inferring,Onnela2021}. 
A customized 5-minute survey was sent to participants every day to inquire about their moods, sleep, social interactions, and psychotic symptoms.

Individualized inference is conducted under the framework of N-of-1 studies, considering the substantial heterogeneity in participants' enrollment time, follow-up duration, negative mood trajectory, and medication use. In Appendix Section G.1 Figure 14, we demonstrate the significant variability in both the trajectory of negative mood (outcome) and estimated coefficients over time for four participants. 
For instance, the estimated coefficients for the outgoing degree of text ($A_{\text{texts},t}$) demonstrate substantial heterogeneity  in terms of significance, magnitude, and direction, making pooled analyses for entire group inappropriate.
For the remainder of the paper, we exemplify our proposed method using a case study of a participant with Bipolar I disorder. Results of other participants are provided in Appendix Section G.1.

\begin{figure}[ht]
\centering
\includegraphics[width=0.96\linewidth]{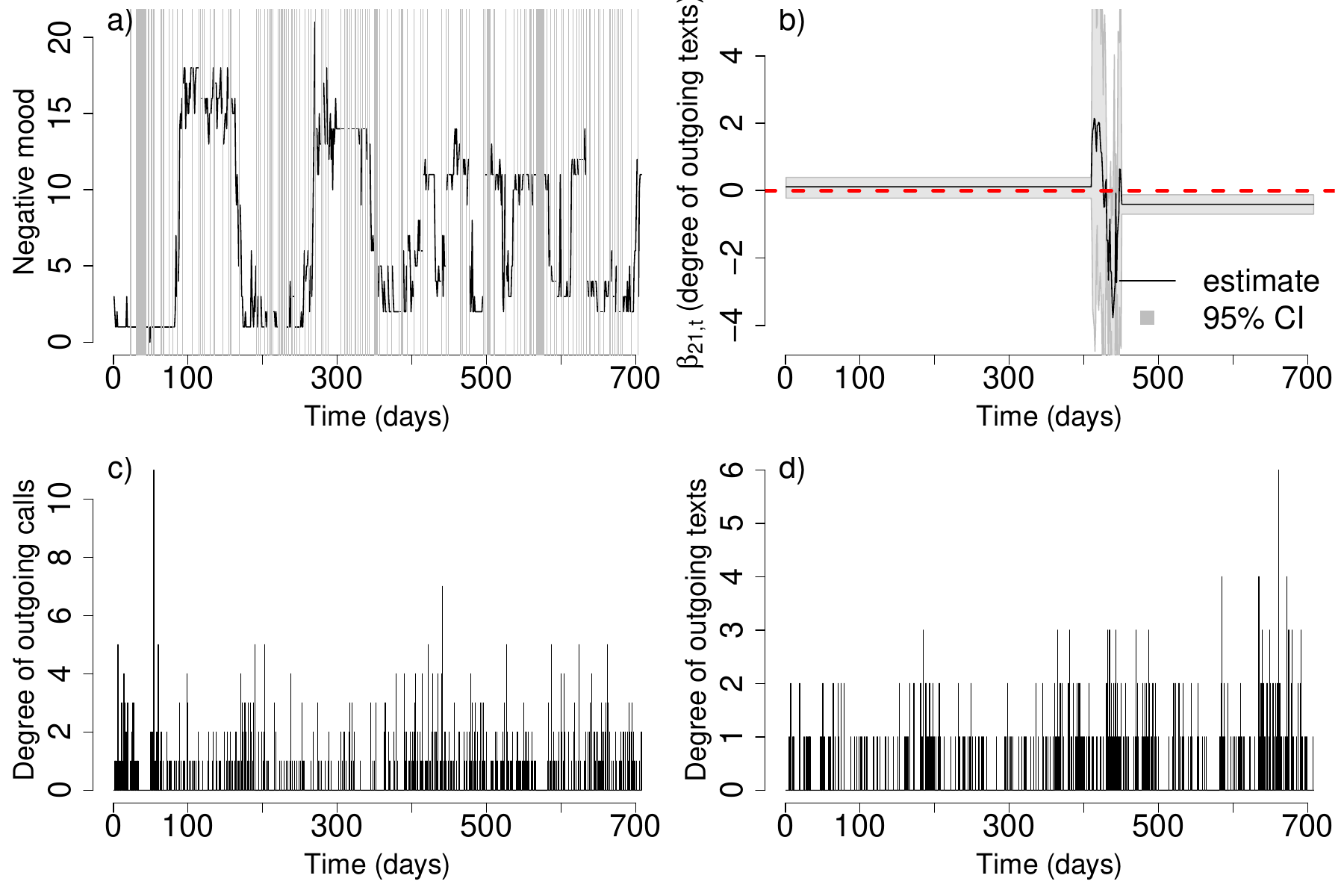}
\caption{Self-reported negative mood (a), estimated $\beta_{21}$ (coefficient for $A_{\text{text,t}}$) (b), degree of outgoing calls (c), and degree of outgoing text (d) over 708 days of follow up for a bipolar patient enrolled in the BLS study. In Figure (a),
gray vertical lines show the days with missing self-reports. In Figure (b), black lines represent the estimated coefficient over time, with $95\%$ confidence intervals represented in grey; the dashed horizontal lines at 0 indicates when there is no effect.  $\beta_{21}$ is found to be periodic-stable with two periods over time with a change point around day 460.}
\label{fig:application}
\end{figure}

Building on prior research, we aim to quantify the effect of phone-based social connectivity on patients’ self-reported negative mood \citep{fowler2022,valeri2023digitalpsychiatry,cai2024causal}. 
In particular, the outcome of interest -- self-reported negative mood ($Y_t$) -- is a composite measure ranging from 0 (best) to 32 (worst), that captures negative feelings experienced by patients comprehensively, including fear, anxiety, embarrassment, hostility, stress, upset, irritated, and loneliness (each ranging from 0 to 4). 
The primary exposures under investigation are phone-based social connectivity features, specifically the degree of contacts via outgoing calls ($A_{calls,t}$) and outgoing texts ($A_{texts,t}$) based on preliminary analysis result \citep{valeri2023digitalpsychiatry}. 
Figure~\ref{fig:application}(a) shows the self-reported negative mood (with 164 days missing), and Figures~\ref{fig:application}(c) and (d) show the passively collected degree of contacts via outgoing calls and texts (with no missing data) during the 708 days follow-up.
Weather temperature ($C_{temp,t}$) \citep{denissen2008effects} and physical mobility ($C_{pm,t}$) \citep{peluso2005physical} have been demonstrated to be associated with negative mood as well as the tendency for social interaction, and are thus included as confounders. 
The physical mobility extracted from accelerometer data are pre-processed following the strategy in \citet{bai2012movelets,bai2014normalization} and aggregated on daily level with details explains in Appendix Section G.2. The trajectory of weather temperature and processed physical mobility data are presented in Appendix Section G.2, Figure 15.


We consider the following state-space model, which accounts for autocorrelation, lagged exposures, and potential confounding:
\vspace{-0.4cm}
\begin{equation}
\begin{split}
Y_t & = \beta_{0,t} + \rho_t Y_{t-1} + \beta_{11,t} A_{calls,t} + \beta_{12,t} A_{calls,t-1} + \beta_{21,t} A_{text,t} + \beta_{22,t} A_{texts,t-1} \\
	& \quad + \beta_{temp,t} C_{temp,t} + \beta_{pm,t} C_{pm,t} + v_t, v_t \sim N(0,R). 
\end{split}
\label{eq:application}
\end{equation}
Additional lagged variables were considered, and model \eqref{eq:application} was selected based on one-step-ahead prediction, a widely used criterion in time series analysis \citep{rivers2002model}. This model is supported by analyses of other participants and aligns with findings from prior research on the BLS study and domain knowledge \citep{fowler2022,valeri2023digitalpsychiatry,cai2024causal}.
Details of the model selection are provided in Appendix Section G.3. Additionally, 
we demonstrate the model's performance by comparing outcome predictions to their observed values along with a $95\%$ CI, as presented in Appendix Section G.5, Figure 17.
For future applications, researchers may consider incorporating additional historical data and confounders as needed.

Table~\ref{tab:analysis_results} presents the results of coefficient estimation using MCEM-SSM, compared to complete case analysis and multiple imputation under state space model specified in \eqref{eq:application}.
 We highlight several key findings from the MCEM-SSM estimation. 
 First, the intercept is modeled as a random walk that effectively captures the drifting nature of negative mood over time. Negative mood exhibits moderate autocorrelation with $\rho_t = 0.42$. 
 Second, the degree of outgoing calls is significantly associated with a decrease in negative mood ($\beta_{11,t}=-0.13$, $\text{CI}=(-0.25,-0.01)$), while the previous day's degree of outgoing calls shows a smaller, non-significant effect ($\beta_{12,t}=-0.02$, $\text{CI}=(-0.15,0.12)$).
Third, the estimated coefficient for degree of outgoing texts is found to be two-period over time (shown in Figure~\ref{fig:application}(b)) with a change point around day 460, prior to which there is no significant association ($\beta_{21,t}(1)=0.11$, $\text{CI}=(-0.23,0.39)$) and subsequent to which we uncover a significant negative association with negative mood ($\beta_{21,t}(2)=-0.41$,  $\text{CI}=(-0.7,-0.13)$). 
This transition occurs concurrently with an increase in Bupropion dosage and a change in psychiatrist. 
The coefficient for degree of outgoing text from the previous day is also significant but smaller in magnitude ($\beta_{22,t}=-0.19$, $\text{CI}=(-0.36,-0.04)$).
In short, the weaker association between previous-day social connectivity and negative mood, compared to that between same-day social connectivity and negative mood, may indicate that social support is more effective on the same day's mood, with its effect decaying the following day.
Finally, two periods are identified for the association between physical mobility and negative mood $\beta_{\text{pm,t}}$ with a change points around day 263, with a significant negative association observed in the second period ($\beta_{\text{pm,t}}(2)=-11.11$, $\text{CI}=(-18.32,-4.90)$), coinciding with the patient's second depressive relapse. No significant association is found between outdoor temperature and negative mood.
Detailed plots of estimated coefficient trajectories over time for all variables are provided in Appendix Section G.4, Figure 16.

The estimation results from complete case analysis (CC) and multiple imputation (mp) are generally consistent with those obtained using MCEM-SSM, except for the coefficients $\beta_{21,t}(2)$ and $\beta_{pm,t}(1)$. These two coefficients are significant using CC and MCEM-SSM but not using mp. 
Additionally, the autocorrelation term $\rho_t$ is overestimated in both CC and MP. This overestimation is further confirmed by simulation results, as shown in Appendix Section F.2, Figure 10.

\section{Discussion}

With the widespread adoption of mobile technology and wearable devices, intensive monitoring is becoming increasingly prevalent in mental health and medical research. 
However, the complexity of multivariate time series data presents significant challenges, particularly in handling missing data. In this study, we propose novel strategies to address missing data problem induced by missing outcomes used as both response and explanatory variables,  in non-stationary multivariate time series arising from observational N-of-1 studies.
Our findings highlight the limitations of commonly used imputation methods, including mean imputation, multiple imputation, last-observation-carried-forward, and ARIMA models, which introduce significant bias in coefficient estimation. 
These methods either fail to generate appropriate imputation candidates for missing values due to their design for static rather than dynamic systems, or they do not adequately account for the entanglement among outcomes, exposures and covariates over time. 
Similarly, complete case analysis introduces bias by selectively omitting missing data and distorting the temporal relationships among variables. 
In contrast, our proposed MCEM-SSM approach preserves the temporal structure of the data, facilitating accurate estimation of time-varying coefficients while incorporating exposure and covariate time series, along with their lagged values, in a dynamic system.

Compared to a fully Bayesian framework, our MCEM-SSM approach is computationally more efficient, leveraging closed-form solutions in optimization for part of parameters. Although MCEM-SSM also requires a moderate convergence time (4 mins 53 seconds), we find it to be a more practical choice for real-world applications involving extended follow-up periods.

Our framework is highly flexible, extending naturally to multivariate time series data from multiple subjects while accounting for trends and seasonality. Under the assumption that subjects follow the same dynamic model over time, multivariate time series from different individuals can be grouped for joint statistical inference. 
In this study, we focused on individualized inference, as the substantial heterogeneity in BLS participants warranted a subject-specific approach. 
However, MCEM-SSM can also accommodate grouped inference, which may be particularly useful in scenarios where participants share similar medical conditions, recruitment timing, and symptom patterns. 
Before applying group-level inference, we recommend conducting individualized analyses to assess the similarity of time series dynamics across participants. If the outcome models demonstrate comparable temporal patterns, group-level inference may enhance statistical power and generalizability. 
On the other hand, if substantial heterogeneity is observed, individualized inference in an N-of-1 setting may be preferred. 
 In short, our proposed method advances the methodology of handling missing data in multivariate non-stationary time series and lays the groundwork for causal inference in non-stationary time series from N-of-1 studies.

BLS is a pioneering effort in long-term smartphone-based monitoring of individuals with severe mental illness (SMI). Using data from a patient with bipolar disorder tracked for nearly two years, we investigated the relationship between phone-based social connectivity and self-reported negative mood. Our model accounts for key confounders, including weather temperature, physical mobility, outcome autocorrelation, and prior exposures of interest.
We identified a moderate autocorrelation in negative mood over time. Additionally, we found a negative association between the degree of outgoing calls and texts and negative mood, and describe how these effects vary over time. 
Critical change points are identified in the effect of degree of outgoing texts, aligning with a psychiatrist change and a significant medication adjustment. 
These findings provide preliminary evidence for the complex role of social network affecting patient's mood over time, potentially depending on symptom severity and environmental factors.

While our MCEM-SSM framework offers improvement in handling missing data in dynamic systems, several limitations need further exploration. 
First, the current implementation of MCEM-SSM assumes linear relationships between outcomes, exposures, covariates, and their lagged values. Future research should explore more flexible models, such as nonlinear state-space frameworks, to account for potential model misspecification.
Additionally, our approach assumes normally distributed outcomes. In the future, we plan to extend it to binary and categorical outcomes.
Furthermore, residual confounding may exist in real-world data analysis as a result of unmeasured individual and contextual confounders. However, the lagged outcomes are likely the strongest confounders in time series analysis, and together with the time-varying intercept, they capture and help control some unmeasured individual and contextual information from the environment. 
As missing data could arise in other predictors (either actively or passively collected), we plan to extend our strategy to address missingness in exposures and confounders in future work. 
Our approach currently assumes missing data mechanisms that are either missing completely at random (MCAR) or missing at random (MAR). Future research should incorporate explicit models to handle MNAR scenarios.
Finally, our analysis reveals significant heterogeneity across participants regarding the association between phone-based social interaction and negative mood. This phenomenon may be attributable to unmeasured confounding variables or latent disease status that modifies the association over time. Future research will investigate how a latent disease modifies the effect of social engagement on mood improvement, providing directions for conducting group inferences for participants with heterogeneous severe mental diseases.

\section*{Acknowledgements}
This work is supported by K01MH118477, U01MH116925. Code used in this paper is available on ``https://github.com/xiaoxuan-cai/MCEM-SSM.git''.
\singlespacing
\bibliographystyle{unsrtnat}
\bibliography{MCEM_SSM.bib}


\begin{footnotesize}
\begin{table}[ht]
\centering
\begin{tabular}{lrrrrrr}
  \hline 
\multirow{2}{*}{Variables} & \multicolumn{2}{c}{Complete Case} & \multicolumn{2}{c}{Multiple Imputation} & \multicolumn{2}{c}{MCEM-SSM} \\ 
\cmidrule(lr){2-3} \cmidrule(lr){4-5} \cmidrule(lr){6-7} 
  & Estimate(SE) & $95\%$ CI & Estimate(SE) & $95\%$ CI & Estimate(SE) & $95\%$ CI\\
  \hline
$\beta_{0,t}$ & \multicolumn{2}{l}{(random walk)} & \multicolumn{2}{l}{(random walk)} & \multicolumn{2}{l}{(random walk)} \\ 
$\rho_{t}$ & 0.71(0.03)** & (0.64,0.78) & 0.84(0.05)** & (0.74,0.94) & 0.42(0.04)** & (0.34,0.5) \\ 
$\beta_{11,t}$ & -0.17(0.08)*\textcolor{white}{*}  & (-0.32,-0.02) & -0.13(0.07)$\dagger$\textcolor{white}{*}  & (-0.26,0.01) & -0.13(0.06)*\textcolor{white}{*}  & (-0.25,-0.01) \\ 
$\beta_{12,t}$ & -0.08(0.07)\textcolor{white}{**} & (-0.22,0.05) & -0.08(0.07)\textcolor{white}{**} & (-0.21,0.06) & -0.02(0.07)\textcolor{white}{**} & (-0.15,0.12) \\ 
$\beta_{21,t}$(1) & 0.04(0.14)\textcolor{white}{**} & (-0.24,0.32) & -0.05(0.15)\textcolor{white}{**} & (-0.35,0.25) & 0.11(0.17)\textcolor{white}{**} & (-0.23,0.39) \\ 
$\beta_{21,t}$(2) & -0.41(0.17)*\textcolor{white}{*}  & (-0.74,-0.08) & -0.23(0.19)\textcolor{white}{**} & (-0.60,0.14) & -0.41(0.15)** & (-0.7,-0.13) \\ 
$\beta_{22,t}$ & -0.20(0.11)$\dagger$\textcolor{white}{*}  & (-0.42,0.02) & -0.23(0.11)*\textcolor{white}{*}  & (-0.45,-0.01) & -0.19(0.09)** & (-0.36,-0.04) \\ 
$\beta_{pm,t}$(1) & -9.43(4.26)*\textcolor{white}{*}  & (-17.78,-1.08) & -4.62(3.27)\textcolor{white}{**} & (-11.02,1.79) & -11.11(3.81)** & (-18.32,-4.90) \\ 
$\beta_{pm,t}$(2) & 1.13(1.6)\textcolor{white}{**}  & (-2.00,4.26) & 0.45(1.21)\textcolor{white}{**} & (-1.92,2.83) & 2.71(1.97)\textcolor{white}{**} & (-0.11,6.88) \\ 
$\beta_{temp,t}$ & 0.00(0.01)\textcolor{white}{**}  & (-0.03,0.02) & 0.00(0.01)\textcolor{white}{**} & (-0.02,0.01) & -0.01(0.01)\textcolor{white}{**} & (-0.04,0.01) \\ 
   \hline
\multicolumn{7}{l}{$\dagger$ $p<0.1$; * $p<0.05$; ** $p<0.01$}
\end{tabular}
\caption{Estimated state-space model coefficients using three data imputation strategies: complete case analysis (left), multiple imputation using MICE (middle), and MCEM-SSM (right). 
Significance level information is indicated by $\dagger$, *, and ** in the footnote. Identified periods for $\beta_{21,t}$ and $\beta_{pm,t}$ are noted in parentheses. All strategies exhibit similar significance patterns.}
\label{tab:analysis_results}
\end{table}
\end{footnotesize}

\end{document}